\documentclass[letter,11pt]{article}

\usepackage{graphicx}
\usepackage{amssymb}
\usepackage{amsmath}
\usepackage{latexsym}
\usepackage{slashed}
\newcommand{\GeV}{~\mathrm{GeV}}
\newcommand{\TeV}{~\mathrm{TeV}}

\makeindex

\title{A REVIEW OF SPIN DETERMINATION AT THE LHC}
\author{Lian-Tao Wang\footnote{lianwang@princeton.edu} and Itay Yavin\footnote{iyavin@princeton.edu} \\
Joseph Henry Laboratories, Princeton University,\\
Princeton, New Jersey 08544, USA \\
}

\begin{document}



\maketitle
\begin{abstract}
Spin measurements are crucial in distinguishing major scenarios of TeV
scale new physics once it is discovered at the LHC. We give a brief
survey of methods of measuring the spin of new physics particles at
the LHC. We focus on the case in which a long lived massive neutral
particle is produced at the end of every cascade decay and escape
detection. This is the case for R-parity preserving supersymmetry,
Little Higgs models with T-parity, extra-dimensional models with
KK-parity, and a large class of similar models and scenarios. After
briefly commenting on measuring spin by combining mass and rate
information, we concentrate on direct measurement by observing angular
correlations among decay products of the new physics particles. We
survey a wide range of possible channels, discuss the construction of
possible correlation variables, and outline experimental
challenges. We also briefly survey the Monte-Carlo tools which are
useful in studying such correlations. 
\end{abstract}


\section{Introduction}

Most models of TeV scale new physics are motivated by solving the
hierarchy problem. Therefore, the most crucial ingredient of all of
them is the mechanism of cancelling the quadratically divergent
correction to the Higgs mass within the Standard Model. In order to
achieve this,  a set of new physics particles with the same or similar
gauge quantum numbers as  
the Standard Model particles are introduced, whose couplings to the
Higgs  are related to those of the Standard Model particles. This
``partner''-like structure of new 
physics is very generic in large classes of new physics
scenarios. Well-known examples include the set of
superpartners in low energy supersymmetry \cite{Dimopoulos:1981zb} (for a recent review see Ref. \cite{Chung:2003fi}) , KK
excitations in 
extra-dimensional models \cite{Appelquist:2000nn}, as well as similar
states in little Higgs models \cite{ArkaniHamed:2001nc} (Ref. \cite{Schmaltz:2005ky} provides a brief review). 

Due to the similarities in gauge quantum numbers, initial LHC signatures
of new partners are very similar, as they can decay into the
same set of observable final state 
particles. The mass spectra of different scenarios can be chosen to
produce similar dominant kinematical features, such as the $p_T$
distribution of the 
decay product. For example, a typical gluino decay chain in
supersymmetry is $\tilde{g} \rightarrow q \bar{q 
} + \tilde{N}_2 $ followed by $\tilde{N}_2 \rightarrow \ell \bar{\ell}
+ \tilde{N}_1$. A similar decay chain in universal extra-dimension
models \cite{Appelquist:2000nn} with KK-gluon ($g^{(1)}$), KK-W ($W^{(1)}_3$)
and KK-photon ($\gamma^{(1)}$), $g^{(1)} \rightarrow q \bar{q} W^{(1)}_3 $
followed by $ W^{(1)}_3 \rightarrow \ell \bar{\ell} \gamma^{(1)}$, gives
identical final states since both $\tilde{N}_1$ and  $\gamma^{1}$ are
neutral stable particles which escape detection. The mass spectra of both
supersymmetry and UED can be adjusted in such a way that the $p_T$ of
the jets and leptons are quite similar. 

Some of the similarities in the LHC signature are actually the result
of equivalences in low energy effective theory. For example, it is
known that ``theory space'' motivated little Higgs models are equivalent to
extra-dimensional models in which 
Higgs is a non-local mode in the bulk, via deconstruction \cite{ArkaniHamed:2001ca,Cheng:2001vd,Cheng:2001nh}. Therefore, they
can actually be described by the same set of low energy ($\sim$ TeV)
degrees of freedom. An important feature of this class of models is
that the partners typically have the same spin as their corresponding
Standard Model particles. 

However, the difference between this set of models and low energy
supersymmetry is physical and observable with 
a sufficiently precise measurement. In particular, the spin of
superpartners differ 
from their Standard Model counter parts by half integers. Therefore,
spin measurements are crucial to set these scenarios apart. 

The conventional way of measuring the spin of a new particle involves
reconstruction of its rest frame using its decay
products and studying the angular distribution about the polarization
axis. For example, in process $e^+ e^- 
\rightarrow Z \rightarrow \mu^+ \mu^- $, the $1+ \cos^2 \theta$
distribution of the muon direction in the rest frame of the $Z$ reveals
its vector nature. However, in most new physics scenarios of
interest such a strategy is complicated by the generic existence of
undetectable massive particles.  Motivated by electroweak precision
constraints and the existence of 
Cold Dark Matter, many such scenarios incorporate some discrete
symmetry which guarantees the existence of a 
lightest stable neutral particle. Well-known examples of such discrete
symmetries include R-parity in 
supersymmetry, KK-parity of universal extra-dimension 
models (UED) \cite{Appelquist:2000nn}, or similarly, T-parity in Little
Higgs Models \cite{Cheng:2003ju,Cheng:2004yc,Low:2004xc,Cheng:2005as}. The 
existence of such a neutral particle at the end of the decay chain
results in large missing energy events in which new 
physics particles are produced. 
This fact helps to separate them from the Standard Model
background. On the other hand, it 
also makes the spin measurement more complicated because it is generically
impossible to reconstruct the momentum, and 
therefore the rest frame, of the decaying new physics particles. 

There are two different approaches to measuring spin. First, given the
same gauge quantum numbers, particles with 
different spin usually have very different production rates, due to the
difference between fermionic and bosonic couplings and the number of degrees
of freedom. Such an approach could be useful, in particular initially, for colored
particles due to their large (hence more measurable) production
rates. However, a crucial ingredient in such a strategy is the measurement of
the masses of particles produced, as rate can only provide definitive
information once the mass is fixed. Such an effort is made more
difficult owing to the existence of missing massive particles. There
is also
some residual model dependence since, for example, a couple of complex
scalars can fake a dirac fermion. 

The second approach,  is the direct measurement of spin through its effect on angular correlations in decay products. In the absence of a
reconstructed rest frame, one is left to consider Lorentz invariant
quantities which encode angular correlations. As we will see later
in this review, spin correlations typically only exist in certain type
of decays. Furthermore, new physics particles are frequently pair
produced with independent decay chains containing similar
particles. Therefore, a valid spin correlation measurement requires
the ability to identify a relatively pure sample of events where we
can isolate certain decay chains and suppress combinatorics
effectively. Therefore, except for very special
cases, we expect this measurement will require large statistics. At
the same time, as will be clear from our discussion, using the appropriate
variables and correctly interpreting the measured angular distribution
frequently requires at least a partial 
knowledge of the spectrum and the gauge quantum numbers. Obtaining
information about the spectrum and the quantum numbers is likely to
require a somewhat lower integrated luminosity than spin measurements
do. Therefore, the order with which we uncover the properties of new
particles is congruent to the order with which we must proceed in the
first place to correctly establish these properties. Thus, we should
clearly focus on mass scales, branching 
ratios and gauge quantum numbers first, once new physics is
discovered at the LHC, while keeping an unbiased perspective towards
the underlying model. More refined measurements, such as the ones
described in this review, will enable us to tell the different models
apart thereafter. Such measurements can be useful and even more powerful in a linear collider as was recently proposed in Ref.~\cite{Buckley:2007th}. In this review we will concentrate on methods applicable to the LHC.

In principle, the production of particles with different spins also leads
to distinguishable angular distributions. This was investigated in the
context of linear colliders in Ref.~\cite{Battaglia:2005zf}. A
similar measurement using the process $pp\rightarrow \tilde{\ell}
\tilde{\ell}^\star$ at the LHC has been studied in
Ref.~\cite{Barr:2005dz}. An analogues measurement in the production
of colored states is more challenging. First, typically several
different initial states and partial waves can contribute to the same
production process. Therefore, it is difficult to extract spin
information from the resulting angular distribution in a model
independent way. Second, as commented above it is often difficult to
reconstruct the direction of the original particles coming out of the
production vertex. As a result, angular correlations are further
washed out.   

In the rest of this review, we will survey both of these approaches
with slightly heavier emphasis given to the angular correlation technique. For
concreteness, we will compare supersymmetry with another generic
scenario in which the partners, such as gluon partner $g'$, W-partner
$W'$, quark partner $q'$, etc., have the same spin as their
corresponding Standard Model particles. As was pointed out above, this
so called same-spin scenario effectively parameterizes almost all
non-SUSY models which address the hierarchy problem. 

Spin measurement at the LHC is still a relatively new field where only
first steps 
towards a comprehensive study have been taken. We will briefly
summarize these developments in this review. We will focus here on the
theoretical foundations and considerations relevant for the
construction of observables. The potential for measuring spin in many
new decay channels remains to be studied.  Important effects, such as
Standard Model background and large combinatorics, deserve careful
further consideration. We outline these issues in connection to
particular channels below.  

\section{Rate and Mass Measurement}

\begin{figure}
\begin{center}
\includegraphics[scale=0.35,angle=270]{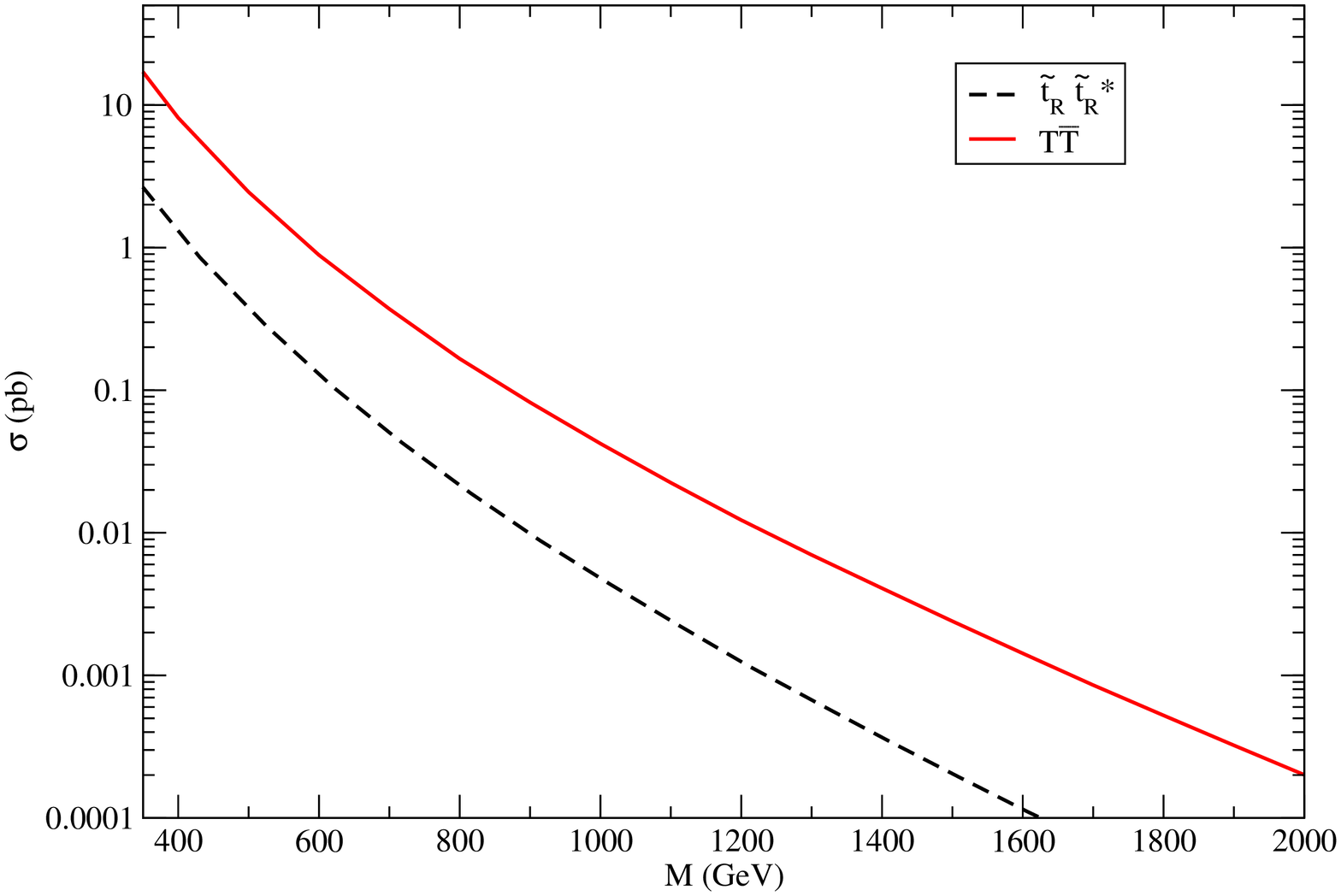}\\
\hspace{-4.5cm}\includegraphics[scale=.9]{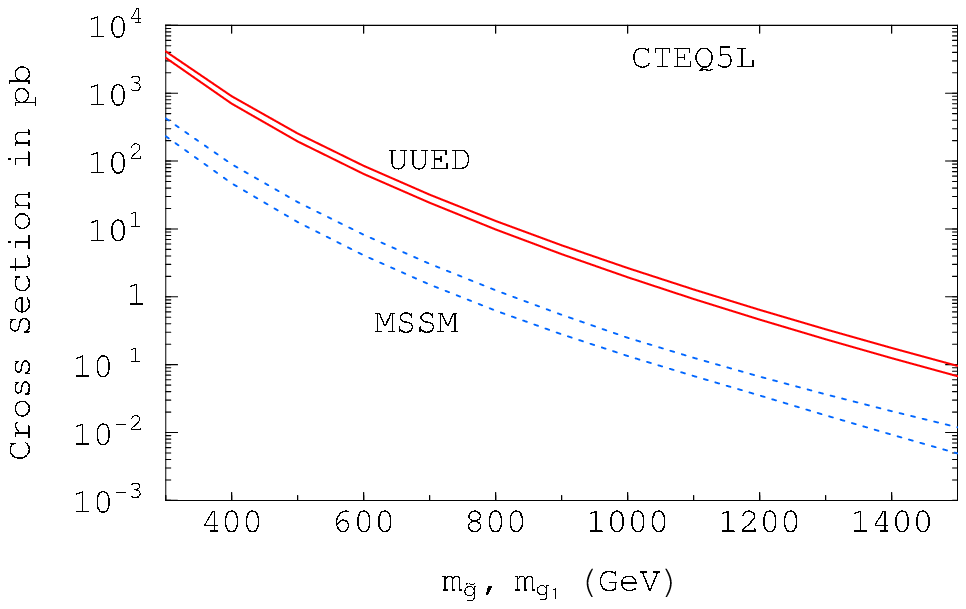}  
\caption{\label{fig:rate} Examples of the production rates of new physics particles. Top
  panel:  the rate of Dirac fermion top partner vs right-handed stop
  (complex scalar) \cite{Cheng:2005as}. Bottom panel: the rate of
  KK-gluon vs gluino (Majorana fermion)\cite{Datta:2005vx}.  }
\end{center}
\end{figure}

The total cross section might serve as an initial hint 
to the spin of the new particles discovered \cite{Datta:2005vx}. Due
to differences in the number of degrees of freedom, the couplings,
spin-statistics and angular momentum conservation, particles with
different spin have very different cross-sections. Two examples of
such differences are shown in Fig.~\ref{fig:rate}. Therefore, if we could
measure the mass of the new physics particles, it will provide us with a
good measurement of the spin. 
Notice that many factors, such as the uncertainties in Parton Distribution
Functions, total integrated luminosity, as well as the  detector and
cut efficiencies have to be taken into account to extract production rates at the LHC. Such difficulties
could in principle be addressed with more 
detailed study and probably higher statistics. A sizable error bar on
the production rate, say a factor of two, will not seriously affect these
results since the rate in different scenarios are very different (see Fig~\ref{fig:rate}) due to the quantized nature of the spin. 

Certain model dependence is inevitable in this approach. For
example, a fermion can be faked by two closely degenerate scalars.
A potentially more serious problem is the model dependence in  extracting
production rates from observations.  What we actually measure is the
production rate multiplied by a some branching
ratio into the particular set of final states we observe. Such
branching ratios could easily vary by one order of magnitude over the
parameter space of a certain scenario. Therefore,
a straightforward extraction of rate information can only be achieved
in the cases where the branching ratio is simple and largely model
independent. This could be possible since there are many
examples in which a single channel dominates a one step decay
process. It is also true when the coupling is simple and known, such
as the decay of some colored particle. However, in more complicated
cases and longer decay chains, more information is necessary to
extract information about the production rate.

Moreover, such a determination is only possible if we could measure
the masses of the particle using kinematical 
information. As demonstrated in \cite{Cheng:2005as,Meade:2006dw},
typical ``transverse'' kinematical observables are mostly only 
sensitive to the mass difference between the decaying particle
and the neutral particle escaping the detector. If this is all the
information one can extract, there are clearly
ambiguities in interpreting data in terms of an underlying model. For
example, suppose supersymmetry is the correct model for TeV scale new physics, and we have observed gluino production  at the LHC, $p p \rightarrow \tilde{g}
\tilde{g} $, followed by $\tilde{g} \rightarrow q \bar{q} + $ LSP.  By assuming supersymmetry, we can find the gluino mass 
$M_{\tilde{g}} $  and LSP mass $M_{\rm LSP}$ which give rise to the
measured rate and mass difference $\Delta m = M_{\tilde{g}} - M_{\rm
  LSP}$. However, there is also, for example, an UED scenario in which
KK-gluon with higher mass  have the same production rate. It follows a
similar decay chain $g^{1} \rightarrow q \bar{q} +$ LKP. We can adjust
the mass of the LKP so that $M_{g^1} - M_{\rm LKP} \sim \Delta m$ at which point this model results in the same experimental observables at the LHC. Notice
that in order for this degeneracies to exist, the only requirement is
that the observables are only functions of $\Delta m$, even if the
functions are different for different scenarios. Of course, we expect
detailed measurement at the LHC to yield more than just the mass
difference. With some assumptions  
regarding the underlying model there are more 
subtle kinematical observables which, in combination with the rate
information, could determine the spin 
\cite{Meade:2006dw}. To what extent this could be generalized to a
broader classes of new physics particles deserves further study. 

Recently, several methods have been proposed to measure the mass of
new physics particles directly\cite{Cheng:2007xv,Cho:2007qv,Gripaios:2007is,Barr:2007hy,Cho:2007dh,Ross:2007rm,Nojiri:2007pq}. A detailed
discussion of these methods is beyond the scope of this
review. Combined with a rate measurement, such determination of mass
will certainly provide an important measurement of the spin within
the context of a certain (possibly well motivated) model. We expect such
information to be complementary to the angular correlations we are
about to discuss next. A detailed comparison between them with the
effect of fully  realistic Standard Model background and detector
effects included has yet to be carried out. 

\section{Angular Correlations in a General Decay Topology}
\label{sec:general}
\begin{figure}[th]
\centerline{\includegraphics[scale=0.8]{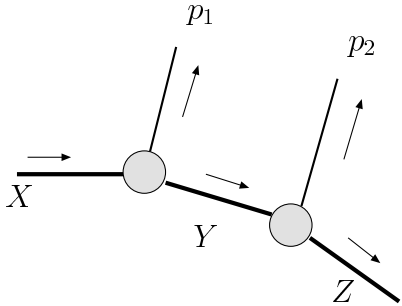}}
\vspace*{8pt}
\caption{The topology for a generic decay where spin information may
be found. A heavy particle $X$ decays into a SM particle with momentum $p_1$
and a heavy partner $Y$ of spin $s$. $Y$ subsequently decays into another SM
particle with momentum $p_2$ and an invisible particle $Z$. If $Y$ is on-shell then the differential decay width for this process is a polynomial of order $2s$ in the relativistically invariant variable $t_{12} = (p_1+p_2)^2$.} 
\protect\label{fig:genericMab}
\end{figure}

The decay topology we will be interested in is shown in
Fig.\ref{fig:genericMab}: A heavy partner, $X$, decays into a SM
particle with momentum $p_1$ and another partner $Y$. $Y$ subsequently
decays into a second SM particle with momentum $p_2$ and a third heavy
partner, $Z$ which may or may not be stable. For most of the
discussion we will assume that $Y$ is produced on-shell, but it is
important to keep in mind that this same topology may still apply even
for 3-body decays where the intermediate particle is off-shell (we
will explore this possibility in Section \ref{sec:off-shell}). If $Y$
is indeed on-shell, it is straightforward to show that the
differential decay width is a polynomial in $t_{12} =
(p_1+p_2)^2$ of degree $2s$, with $s$ being the spin of $Y$ (for
details see \cite{Wang:2006hk}), 
\begin{equation}
\label{eqn:dGamma}
\frac{d\Gamma_{X\rightarrow p_1~p_2~Z}}{dt_{12}} = a_0 + \ldots + a_{2s} t_{12}^{2s} \quad \quad \text{where} \quad t_{12} = (p_1 + p_2)^2
\end{equation} 
We have chosen to use the variable $t_{12} = m_{12}^2$, rather than invariant mass $m_{12}$. While they are equivalent in principle, distribution of $t_{12}$ is more convenient to interpret since it is linear in $\cos \theta_{12}$. 

The observed distribution often deviates from this simple relation because of both experimental and theoretical reasons.  First, if the intermediate particle $Y$ is not polarized, there is no hope of seeing any angular correlations, i.e. the polynomial is simply a constant. Second, if it is a fermion then polarizing it is not enough and we need its decay to distinguish handedness. This leads us to some simple necessary conditions for the existence of angular correlations\cite{Wang:2006hk},
\begin{enumerate}
 \item If the intermediate particle $Y$ is a vector-boson then $M_Y \ll M_X$ so that it is longitudinally dominated and hence polarized.
 \item If the intermediate particle is a Dirac fermion then both
   vertices in the topology of Fig. \ref{fig:genericMab} must be at
   least partially chiral\footnote{We will see below that there is
     possibly an additional subtle kinematical suppression if the
     fermion consequently decays into a heavy vector boson, but this
     is non-generic.}.  
 \item If the intermediate particle is a Majorana fermion then one must be able to determine the charge of the outgoing particles, $p_1$ and $p_2$ (even if only statistically). 
\end{enumerate}

Based on these simple rules it is straightforward to list the
different decay channels that may contain spin effects in the MSSM and
other scenarios. In what follows, we divide these channels into those
which probe the spin of the Electroweak gauge-bosons and Higgs sector
partners and those which probe the spin of the matter sector
partners. In certain cases additional information of the spin
correlations can be used to determine the spin of the external
particles, $X$ and $Z$ in Fig. \ref{fig:genericMab}. 
It can lead to a full spin determination of every new physics particle
involved in a decay chain.  

\section{Mis-pairing and Background}

\begin{figure}[th]
\centerline{\includegraphics[scale=0.6]{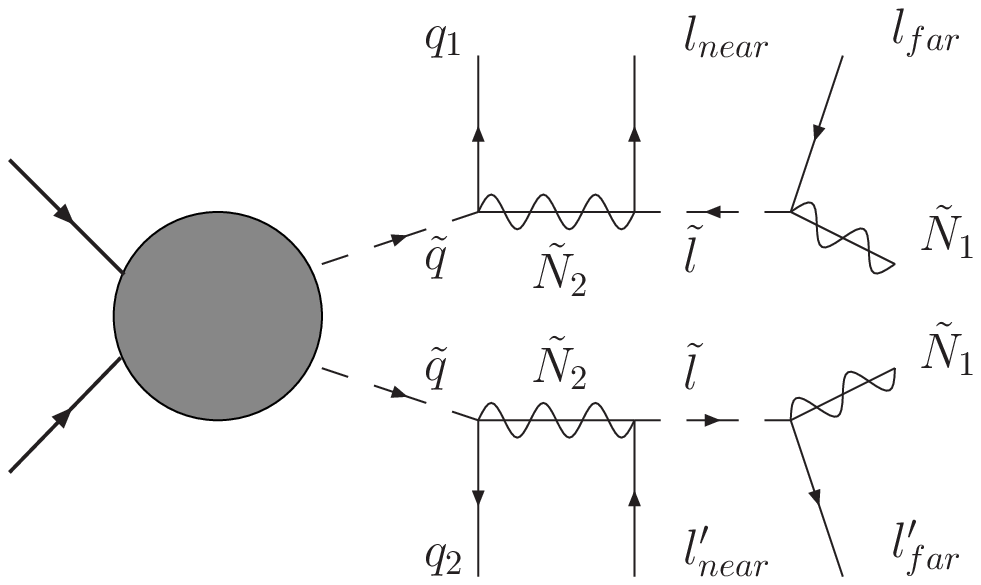}}
\vspace*{8pt}
\caption{A typical SUSY decay chain where the identification of the right pairing required for spin determination could be difficult. } 
\protect\label{fig:ComboProb}
\end{figure}

Common to all attempts of establishing angular correlations is the
problem of mis-pairing. The generic process shown in
Fig. \ref{fig:genericMab} is usually a part of a longer decay chain
involving more final state particles. Moreover, this decay chain
itself is only one of two branches in pair production. For example,
the full event may look like Fig. \ref{fig:ComboProb} , which is a
process commonly discussed in connection with spin determination and
elaborated upon in Section (\ref{sec:EWinos}). If we are interested in
the spin of $\tilde{N}_2$ we would like to correlate the adjacent
outgoing jet and lepton, e.g. $q_1$ and $l_{near}$. However, in
general there are no sharp kinematical differences between the near
and far leptons along the same decay chain. Similarly, it is not easy
to tell which is the correct jet to pair with the lepton. Therefore,
an experimental histogram of the invariant mass variable $t_{12}$ is
unlikely to follow the simple polynomial presented of
Eq. (\ref{eqn:dGamma}). Instead we will get some shape which is
composed of the polynomial structure coming from the correct pairing
together with a more featureless distribution associated with the wrong
pairing. This is in principle not a problem since one can fit to an
underlying model by simulation, however, it does lead to a reduction in the
relative statistical significance of the correct pairing. This problem
is shared by all the channels we 
discuss below and we will refrain from the refrain where it is
self-evident.  

We will not discuss the problems associated with Standard Model background in this review and assume that it was reduced with a strong set of cuts on missing energy, number of leptons and etc. However, even if all the events contain new physics we may still have events with different underlying topology contributing the same visible final states. We refer to such ambiguities as same model background. Before investigating spin effects we must first ascertain that the event topology is well understood. Otherwise the resulting correlations will carry little significance. 

\section{Spin Determination of Electroweak Gauge-Boson Partners}
\label{sec:EWinos}
In this section we cover potential channels which may reveal the spin of heavy partners of the EW gauge-bosons and scalars (in SUSY those would be the winos, binos and higgsinos). Considering the hierarchy problem it is necessary to have the $W^\pm$ gauge-boson partners at the EW scale since the $W^\pm$ gauge-boson loop contribution to the corrections to the Higgs mass are the largest after the top quark loops. We begin with the least model dependent channel and follow with channels which involve more assumptions about the spectrum. 

\subsection{Charged boson partner's spin - Jet-$W^\pm$ correlations}
\begin{figure}[ht]
\begin{center}
\includegraphics[scale=0.7]{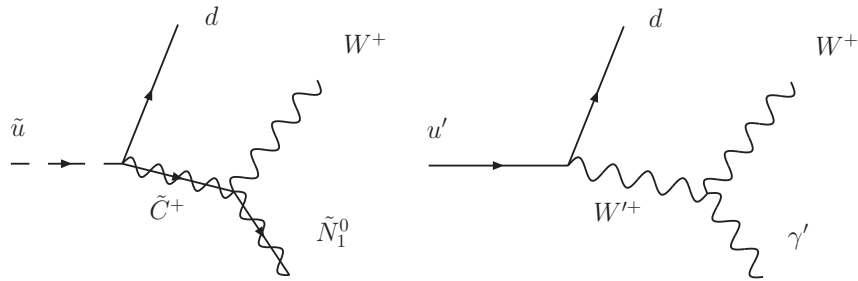}
\caption{Topology for the decay of an up-type quark partner into a down-type quark, a $W^+$ gauge-boson and an invisible stable particle. On the left is the SUSY decay chain and on the right is the corresponding decay chain in a theory with same-spin partners. } 
\label{fig:qWchannel}
\end{center}
\end{figure}

The first channel we discuss is the minimal one in that it only assumes the presence of quark partners, charged EW bosons' partners and a stable lightest particle. This channel was first presented and investigated in Ref.~\cite{Wang:2006hk} and later also in Ref.~\cite{Smillie:2006cd}. The relevant topology is shown in Fig. \ref{fig:qWchannel} for both SUSY and same-spin scenarios. 

The differential decay width for the two processes is a polynomial in $t_{qW} = (p_q + p_W)^2$, where $p_q$ is the momentum of the outgoing quark and $p_W$ is that of the outgoing $W$ gauge-boson. It is of degree 1 (2) for the SUSY (same-spin) scenario and is given by,
\begin{equation}
\label{eqn:qWchannel}
\frac{d \Gamma_{SUSY}}{dt_{qW}} = C_1 (|a_L|^2 - |a_R|^2) t_{qW} + C_0 \quad \quad\quad\frac{d \Gamma_{same-spin}}{dt_{qW}}  = C_2^\prime t_{qW}^2 + C_1^\prime t_{qW} + C_0^\prime
\end{equation}
where $a_{L,R}$ are the left and right coupling of the $\tilde{N}_1^0-\tilde{C}^+-W^+$ vertex and the $C_i$'s are dimensionful factors which depend on the masses of the different particles in the decay and the quark partner's coupling. The leading coefficients are given by,
\begin{eqnarray}
C_1 &=& \frac{g_2^2}{512\pi^2}\frac{M_{\tilde{C}}\left(M_{\tilde{C}}^2-2M_{W}^2+M_{\tilde{N}}^2\right)}{M_{\tilde{q}}^3 M_W^2 \Gamma_{\tilde{C}}}  \\
C_2^\prime &=& \frac{g_2^2}{256\pi^2}\frac{ M_W^4  + \left(M_{W^\prime}^2-M_{\gamma^\prime}^2\right)^2 + 2M_W^2\left(5M_{\gamma^\prime}^2-M_{W^\prime}^2\right)}{ M_{q^\prime}^3 M_W^2 M_{\gamma^\prime}^2 M_{W^\prime}\Gamma_{W^\prime}}
\end{eqnarray}
where $\Gamma_{\tilde{C}}$ ($\Gamma_{W^\prime}$) is the decay-width of the chargino ($W^\prime$) and we assumed that the quark partner coupling is dominated by its weak interaction. The expression for $d \Gamma_{SUSY}/dt_{qW}$ in Eq.(\ref{eqn:qWchannel}) makes it evident that in the SUSY case angular correlations vanish in the non-chiral limit where $|a_L| = |a_R|$. 

There are several experimental difficulties associated with this channel. First, if the $W^\pm$ decays leptonically then one must correlate the resulting lepton with the outgoing jet. Since it is impossible to fully reconstruct the gauge-boson the distribution is no longer governed by the simple formula in Eq.(\ref{eqn:qWchannel}) (if the gauge-boson is strongly boosted in the lab frame one may be able to extract its four-vector at least approximately). Nonetheless, preliminary studies seem to indicate that the distinction between SUSY and same-spin scenarios is still very pronounced\cite{Wang:2006hk}. 

Second, if the $W^\pm$ decays hadronically then it is important to
establish how well can the $W^\pm$ be reconstructed and its momentum
correlated with that of the outgoing jet. This will require a more
realistic study and simulations.

\begin{figure}[th]
\begin{center}
\includegraphics[scale=.4,angle=270]{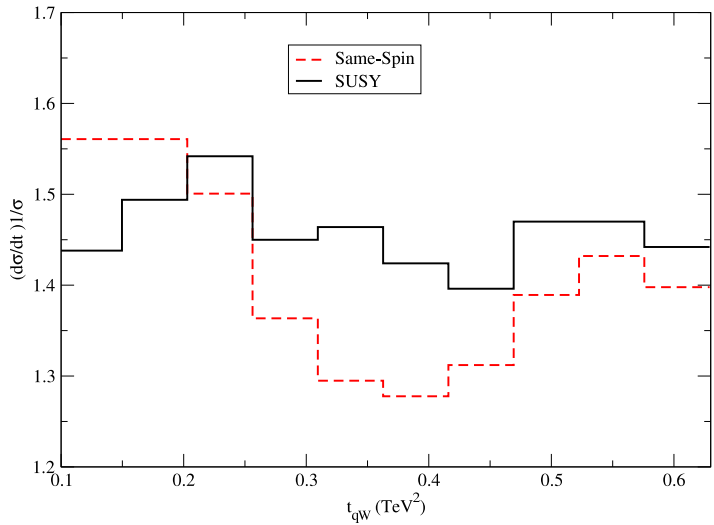}
\end{center}
\caption{Considering only signal events with 2-jets and a hadronic $W^\pm$ this is a histogram of the invariant mass $t_{qW}$ (see text for details concerning the reconstruction) for the two prototype models: SUSY (solid-black), Same-Spin (dashed-red). The histogram is normalized to unit area and the normalized statistical error is approximately $\sqrt{N}\sim 0.04$. The quadratic behavior of the same-spin scenario is clearly distinguished from the linear behavior associated with SUSY.} 
\label{fig:JetsWmc}
\end{figure}

Fig.(\ref{fig:JetsWmc}) contrasts the Jet-$W^\pm$ correlation in SUSY against same-spin scenarios with quark partners at $1\TeV$, vector boson partners at $500\GeV$ and an LSP at $100\GeV$. This simulation was done including background from the same model (no SM background, though), i.e. events with identical final states, but different underlaying topology and without any knowledge of the correct pairing. In every event one tries to reconstruct the $W^\pm$ from two of the 4 jets and then form the invariant mass $t_{qW}$ with the two remaining jets. If more than one pairing reconstructs $W$ it is regarded as failure and the event is discarded. The cuts involve $\slashed{P}_T > 200\GeV$ and $|\eta| <4.0$. Jets were defined with a cone size of $\Delta R = 0.4$. The linear behavior vs. the the quadratic behavior is still visible on top of the background coming from the wrong pairing and contamination from other event topologies.

This example also serves as a remainder that any observable suggested as a possible determinant of spin better result in robust differences in distributions' shapes. Any discriminator that relies on small discrepancies is unlikely to survive when the experimental limitations are properly taken into account.

\subsection{Charged boson partner's spin - Jet-$Z^0$ correlations}

\begin{figure}[th]
\begin{center}
\includegraphics[scale=0.7]{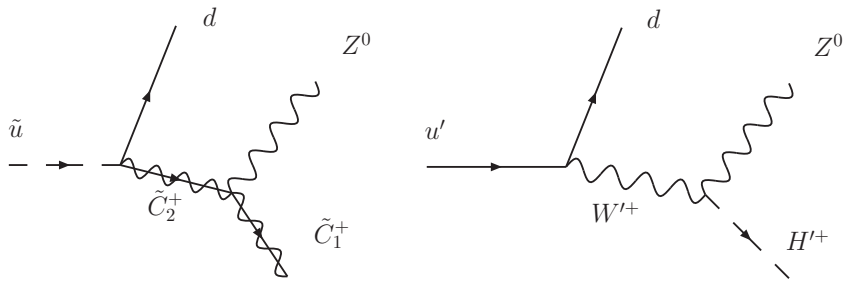}
\caption{Topology for the decay of an up-type quark partner into an up-type quark, a $Z^0$ gauge-boson and a charged partner. On the left is the SUSY decay chain and on the right is the corresponding decay chain in a theory with same-spin partners. } 
\label{fig:qZchannel}
\end{center}
\end{figure}

If the spectrum contains two light charged EW bosons' partners
(e.g. both Higgsino and Wino are light) then the topology shown in
Fig. \ref{fig:qZchannel} is possible. With sizable branching ratio,
this is a very promising decay channel because the four-momentum of
the $Z^0$ gauge-boson can be unambiguously reconstructed. The
differential decay width follows that of 
Eq. (\ref{eqn:qWchannel}) and the prospects of determining spin using
this channel are now under investigation\cite{Hook:2008}. In addition,
if the underlying model is indeed SUSY, the slope in
Eq. (\ref{eqn:qWchannel}) is sensitive to the value of
$\tan\beta$. This is easy to understand if one recalls that for an
intermediate Dirac fermion both interaction vertices need to be at
least partially chiral. The $\tilde{q}-\tilde{C}_2-q$ vertex is
certainly chiral if the intermediate chargino is mostly a wino. The
$\tilde{C}_2-\tilde{C}_1-Z$, however, is only chiral if $\tan\beta \ne
1$. In fact it is straightforward to show that the slope of the
distribution is directly related to $\tan\beta$ and vanishes as
$\tan\beta \rightarrow 1$. Since, this measurement is in principle
very clean, it may offer extra information in addition to the spin of
the chargino state.

\subsection{Neutral boson partner's spin}

\begin{figure}[ht]
\begin{center}
\includegraphics[scale=0.7]{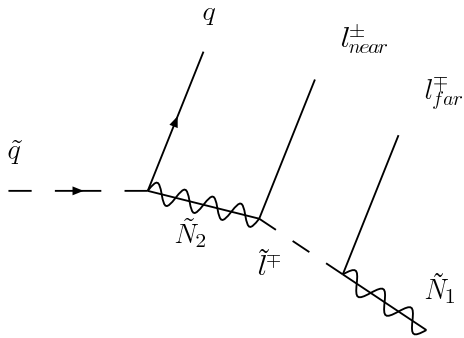}
\caption{Topology for the cascade decay of a quark into two leptons and neutral stable partner. The correlation of the quark with the near lepton reveals the spin of the intermediate neutralino $\tilde{N}_1$. However, experimentally it is hard to distinguish between the far and near lepton and one must average over both. The resulting distribution is asymmetric in the correlation of the quark with a positive versus negative leptons.} 
\label{fig:qllchannel}
\end{center}
\end{figure}

In an early and interesting paper \cite{Barr:2004ze}, Barr has pointed out that the inherent charge asymmetry of a proton-proton collider may lead to certain angular correlations which carry spin information. This channel is known to be useful for mass determination and its sensitivity to spin was pointed out earlier in Ref.~\cite{Richardson:2001df} and references therein. The decay topology for the SUSY case is shown in Fig. \ref{fig:qllchannel}. 

In the case of SUSY, the intermmediate particle is a Majorana fermion and therefore, as pointed above, the charge of the outgoing quark and the near lepton must be determined in order to observe spin effects. The quark's charge may be obtained by either utilizing the charge asymmetry of the proton as pointed out by Barr, or by considering decays involving heavy flavor quarks such as bottom-quarks where a statistical determination of the charge is possible \cite{Alves:2006df}. The lepton's charge can be determined unambigiously, but experimentally, the far lepton cannot be distinguished from the near lepton and one must average over both contributions. Barr found that it is possible to define an asymmetry parameter which is sensitive to the spin of the intermediate neutralino $\tilde{N}_2$,
\begin{equation}
\label{eqn:BarrAsym}
\mathcal{A} = \left( \frac{d\Gamma}{dt_{ql^+}} -\frac{d\Gamma}{dt_{ql^-}}\right)\biggr/\left(\frac{d\Gamma}{dt_{ql^+}} + \frac{d\Gamma}{dt_{ql^-} }\right) 
\end{equation}
(The original asymmetry defined by Barr was with respect to the invariant mass rather than $t_{ql^\pm}$, but these two definition are isomorphic). As in the previous cases, this asymmetry has the opposite sign if the initial state is an anti-squark. Therefore, this observable relies on the LHC being a proton-proton collider and is sensitive to the quark - anti-quark asymmetries in the parton distribution functions.

This decay chain was further investigated by \cite{Battaglia:2005zf,Smillie:2005ar,Datta:2005zs,Athanasiou:2006ef} and contrasted against the corresponding decay chain in the scenario with Universal Extra Dimensions (UED) \cite{Appelquist:2000nn,Cheng:2002ab}. It is worthwhile to point out that when the intermmediate particle is a vector-boson (as in UED), no charge information is necessary in order to observe spin effects. However, correlations are washed out when the spectrum becomes degenerate.   

We remark that this channel demands the existence of light lepton partners in the spectrum. Leptonic partners at the EW scale are certainly not required in order to solve the hierarchy problem. For example, they are not present in Little Higgs models. They exist in Supersymmetric theories and RGE effects tend to cause them to be lighter than the quark partners, but they may not necessarily be lighter than the neutralinos. Nonetheless, if they are light and the decay chain of Fig. (\ref{fig:qllchannel}) is allowed, it will teach us an enormous amount regarding the underlying model. 

Some of the experimental difficulties associated with this channel
were pointed out above already (mis-pairing and same model
background). In addition, the necessity for a charge asymmetry means
that gluino pair production events will tend to contaminate the signal
and reduce the sensitivity due to the absence of charge asymmetry to
begin with. This is also true for squark pair production if dominated
by gluon-gluon initial state. In connection with this difficulty,
Ref.~\cite{Alves:2006df} had the interesting proposal of using the
distinction between $b$ and $\bar{b}$. It may even lead to a
measurement of the gluino spin with enough luminosity. However, so far
this possibility has only been investigated for the parameter point
SPS1a which is particularly friendly for hadron collider studies. It
remains uncertain whether these conclusions can be extended to more
generic spectra.  

\section{Spin Determination of Standard Model Matter Partners}
\label{sec:Smatter} 

In SUSY, the matter partners are scalars and therefore carry no spin information. In scenarios with same spin partners, angular correlations in decays involving the intermediate heavy fermions are certainly possible and the conditions for such effects were elucidated in Ref.~\cite{Kilic:2007zk}. It was also noted that, in combination with the measurements discussed above, one may obtain information on the spin of the initial and final partners in the cascade. We now briefly review these findings.

\begin{figure}[ht]
\begin{center}
\includegraphics[scale=0.7]{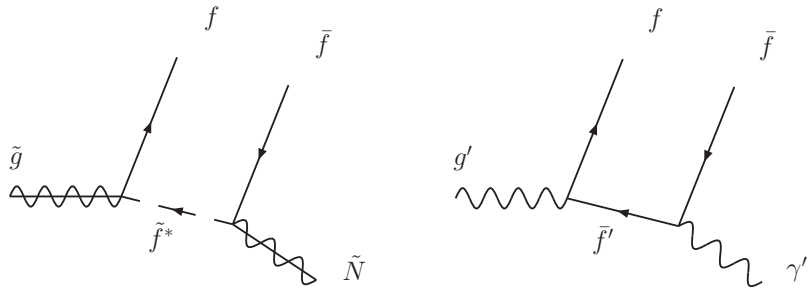}
\caption{Topology for a decay involving an intermediate matter partner. On the left is the SUSY decay of a gluino into two SM fermions and a neutralino. Since the intermediate particle is a scalar there are no angular correlations between the two outgoing fermions. On the right is the decay in a scenario with same spin partners. A heavy gluon-like particle decays into two SM fermions and a heavy neutral gauge-boson. The outgoing fermions do exhibit angular correlations owing to the fermionic nature of the intermediate heavy matter partner.} 
\label{fig:intFchannel}
\end{center}
\end{figure}

The general decay topology is shown in Fig.\ref{fig:intFchannel} where the SUSY decay of a gluino is shown alongside the corresponding decay in the same-spin scenario. The differential decay width is in general given by,
\begin{equation}
\label{eqn:intFchannel}
\frac{d \Gamma_{SUSY}}{dt_{f\bar{f}}} = C_0 \quad \quad\quad\quad\frac{d \Gamma_{same-spin}}{dt_{f\bar{f}}}  =  C_1^\prime t_{f\bar{f}} + C_0^\prime
\end{equation}
where $t_{f\bar{f}} = (p_f + p_{\bar{f}})^2$ and the coefficients $C_1^\prime$ and $C_0^\prime$ are presented in Table \ref{tbl:diffspin}.

As mentioned above, when the intermediate partner is a Dirac fermion the necessary conditions for the existence of angular correlations (i.e. for the coefficient $C_1^\prime$ in Eq.(\ref{eqn:intFchannel}) to be non-zero) are that both interaction vertices in Fig.\ref{fig:intFchannel} are at least partially chiral. As we will now show, it is possible to satisfy this condition with very few assumptions about the same-spin model. 

First, we will assume that the members of this new sector have the same quantum numbers as the known quarks and leptons (this assumption is not necessary, but will simplify the discussion). As throughout this review,  we will assume that some $Z_2$ parity symmetry is present. In what follows we use the quark sector to demonstrate our results, but similar conclusions hold for the lepton sector as well.

\begin{table}[ht]
\begin{center}
\begin{tabular}{c|c|c}
SM & Heavy partners  &$SU(2)\times U_Y(1)$  \\
\hline $q_L$ & \parbox{6cm}{$Q_L^\prime = \left(
\begin{array}{l}
u_L^\prime \\
d_L^\prime \\
\end{array}
\right) \quad Q_R^\prime = \left(
\begin{array}{l}
u_R^\prime \\
d_R^\prime \\
\end{array}
\right)$} & $\left(2,\frac{1}{6}\right)$
\\
$u_R$ & $U_R^\prime \hspace{2.2cm} U_L^\prime$ &
$\left(1,\frac{2}{3}\right)$
\\
$d_R$ & $D_R^\prime \hspace{2.2cm} D_L^\prime$ &
$\left(1,-\frac{1}{3}\right)$
\end{tabular}
\caption{The matter content for the heavy partners in a generic same-spin scenario.}
\label{tbl:matter}
\end{center}
\end{table}
We denote the heavy fermionic modes by $Q_{L,R}^\prime$, $U_{L,R}^\prime$ and $D_{L,R}^\prime$ as shown in Table \ref{tbl:matter}. Here, $Q^\prime = (u^\prime,d^\prime)$ is an $SU(2)$ doublet while $U^\prime$ and $D^\prime$ are singlets; note that $Q_L^\prime$, $U_R^\prime$ and $D_R^\prime$ are partners to SM fermions while $Q_R^\prime$, $U_L^\prime$ and $D_L^\prime$ have no Standard Model counterparts.

At this stage, the coupling to the SM matter is via heavy bosons ($g^\prime$, $W^\prime$ and etc.) because of the $Z_2$ parity. Since the SM matter sector is chiral it forces the new interaction to be chiral as well. The coupling to the heavy gluon $g^\prime$, for example, is schematically,
\begin{equation}
\label{eqn:interactions}
\mathcal{L}_{int} = \overline{Q}^\prime_L \slashed{g}^\prime q_L +
\overline{U}^\prime_R \slashed{g}^\prime u_R + \overline{D}^\prime_R \slashed{g}^\prime
d_R + h.c.
\end{equation}
where $q_L$ is the SM electroweak doublet and $u_R$ and $d_R$ are the singlets.

However, for the new fermions to be parametrically heavy they must have Dirac masses in addition to the usual Yukawa couplings to the Higgs. Therefore, after EW symmetry breaking, their mass matrix is given by (in here we ignore the flavor structure to simplify the discussion),
\begin{equation}
\mathcal{L}_{mass,up} = \left(
\begin{array}{cc}
u^\prime_L & U^\prime_L
\end{array}
\right)
\left(
\begin{array}{cc}
M_Q & \lambda v \\
\lambda v & M_U~e^{i\varphi}
\end{array}
\right)
\left(
\begin{array}{c}
u^\prime_R \\
U^\prime_R
\end{array}
\right).
\label{eqn:massmatrix}
\end{equation}
where $v=246\GeV$, $\lambda$ is a Yukawa coupling of the new states, and $\varphi$ is some phase which in general cannot be rotated away. A similar mass matrix holds for the down sector. After diagonalizing the mass matrix the mixing angle is given by,
\begin{equation}
\tan{\theta}=\frac{2\lambda v \sqrt{M_Q^2 + 2M_QM_U \cos\varphi + M_U^2}}{\left(M_Q^2-M_U^2\right) + \sqrt{\left(M_Q^2-M_U^2\right)^2 + 4\lambda^2v^2\left(M_Q^2 + 2M_QM_U\cos\varphi + M_U^2\right)}}
\label{eqn:rotationangles}
\end{equation}
Below, we concentrate on two limiting cases of this relation which result in very different angular correlations. 

\subsection{Non-degenerate Spectrum}

When the diagonal terms in the mass matrix are sufficiently separated, $M_Q - M_U \gg \lambda v$, the mixing between the states is small and the phase plays no role,
\begin{equation}
\label{eqn:non-deg-mix}
\tan{\theta} \sim \frac{\lambda v}{\Delta M} \quad \ll \quad 1
\end{equation}
where $\Delta M = M_Q - M_U$. When the mixing is small the interactions with the SM in Eq.(\ref{eqn:interactions}) remain chiral after rotation into the mass eigenstate basis. This leads to an important conclusion: any model with heavy
fermionic partners of the SM matter sector protected by some $Z_2$ parity ($KK$-parity, $T$-parity etc.), with a non-degenerate
spectrum, exhibits angular correlations in decays.

\subsection{Degenerate Spectrum}

We now turn to examine a degenerate spectrum, so that $M_Q - M_U \ll \lambda v$. Such a spectrum can result from any symmetry that
relates left and right mass parameters. In this case the phase $\varphi$ carries significance. When $\varphi=0$ the mixing is large,
\begin{equation}
\tan{\theta} = \frac{\lambda v}{\Delta M + \sqrt{\Delta M^2 + \lambda^2 v^2}} \rightarrow 1 - \frac{\Delta M}{\lambda v}
\end{equation}
The coupling of the mass eigenstates to the SM is no longer chiral,
\begin{equation}
\label{eqn:nonChiralInt}
\mathcal{L}_{int} = \overline{U}_1^\prime \slashed{g}^\prime u + \overline{U}_2^\prime \gamma_5 \slashed{g}^\prime u \quad +\quad  \mathcal{O}\left(\frac{\Delta M}{\lambda v}\right) + h.c.
\end{equation}
and similarly for the other gauge couplings. Therefore, since the interactions with the SM, Eq.(\ref{eqn:nonChiralInt}), are no longer chiral we expect angular correlations in decays to vanish (up to corrections of order $\mathcal{O}(\Delta M/\lambda v)$).

In the UED model, not only are the masses degenerate by construction, but also the phase is fixed by 5-d Lorentz invariance
$\varphi = \pi$ as shown in Ref.~\cite{Kilic:2007zk}. In this case, unless $\lambda$ is unnaturally large, the mixing is always small,
\begin{equation}
\tan{\theta} = \frac{\lambda v}{\overline{M} +
\sqrt{\overline{M}^2+v^2\lambda^2}} \rightarrow \frac{\lambda
v}{2\overline{M}} \quad \ll \quad 1
\end{equation}
where $\overline{M} = (M_Q+M_U)/2$. Therefore, as can be seen from
Eq.(\ref{eqn:interactions}), the interactions of the SM with 
$U_1^\prime$ and $U_2^\prime$ remain chiral and angular correlations
should be present in decays.  

This observation leads to the following conclusion in the case of
degenerate spectrum: \textit{if angular
  correlations are not present, or are strongly suppressed, UED model
  can be ruled out. At the same time, it is still possible that matter sector
  partners are fermions.} 

\subsection{Slope Information}

If the vertices are chiral and angular correlations are therefore
present then one may extract more information out of such measurements
than just the spin of the intermediate matter partner. In Table
\ref{tbl:diffspin} the other possible spin assignments for the
external particles in a decay involving an intermediate Dirac fermion
are shown\footnote{The reader will undoubtedly notice that the slope
  vanishes as $M_Q^2 \rightarrow 2 M_{\gamma^\prime}^2$ when
  $\gamma^\prime$ is a vector-boson. This subtle kinematic suppression
  arises because the outgoing vector-boson $\gamma^\prime$ is
  unpolarized in this limit.}. All these distributions have an edge
at, 
\begin{equation}
t_{f\bar{f}}^{(edge)} = \frac{\left(M_{g^\prime}^2 - M_{Q}^2 \right)  \left(M_{Q}^2 - M_{\gamma^\prime}^2 \right)}{M_Q^2}
\end{equation}
However, depending on the spin of the external particles the slope is
different and the behavior near the edge is modified. The slopes
presented in the table assume no significant Left-Right mixing, such
as UED or the non-degenerate case discussed above. A determination of
the slope together with a measurement of the ratio
$M_Q/M_{\gamma^\prime}$ (from kinematical edges and cross-sections)
can identify the spin of the external particle up to a two fold
ambiguity. 

For example, if one measures a positive slope and $M_Q^2/M_{\gamma^\prime}^2 > 2$ then one can conclude that the gluon partner
is a vector-boson, but whether $\gamma^\prime$ is a scalar or a vector-boson is left unresolved. On the other hand, a positive slope together with $M_Q^2/M_{\gamma^\prime}^2 > 2$ leads one to conclude that $\gamma^\prime$ is a scalar, but the spin of the gluon partner can be either zero or unity. 

\begin{table}[ht]
\begin{center}
{
\begin{tabular}{c|c|c}
Scenario & Slope  $C_1^\prime$ & Intercept $C_0^\prime$ \\
\hline
\includegraphics{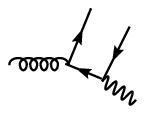} & $\left(2 M_{g^\prime}^2- M_Q^2 \right)\left(M_Q^2 - 2M_{\gamma^\prime}^2 \right)$ & $(M_{Q}^4 + 4 M_{\gamma^\prime}^2 M_{g^\prime}^2) ~t_{f\bar{f}}^{(edge)}$\\
\includegraphics{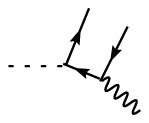} & $-\left(M_Q^2 - 2M_{\gamma^\prime}^2 \right)$ & $ M_Q^2 ~ t_{f\bar{f}}^{(edge)} $\\
\includegraphics{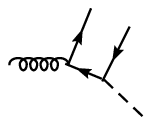} & $\left(2M_{g^\prime}^2 - M_Q^2 \right)$ & $M_Q^2 ~ t_{f\bar{f}}^{(edge)} $\\
\includegraphics{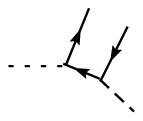} & $-1$ & $ t_{f\bar{f}}^{(edge)} $\\
\end{tabular}
}
\caption{If angular correlations exist between the outgoing
$f-\bar{f}$ or dilepton pair, then the sign of the slope of the
distribution (whether $C_1^\prime>0$ or $C_1^\prime <0$) may reveal the spin of the external particles
as well as the intermediate one. In the first row we consider a
scenario where the external particles are both vector-bosons (VFV). In the
second row the incoming partner is a scalar whereas the outgoing
partner is a vector-boson (SFV) and so forth.}
\label{tbl:diffspin}
\end{center}
\end{table}

\begin{figure}[ht]
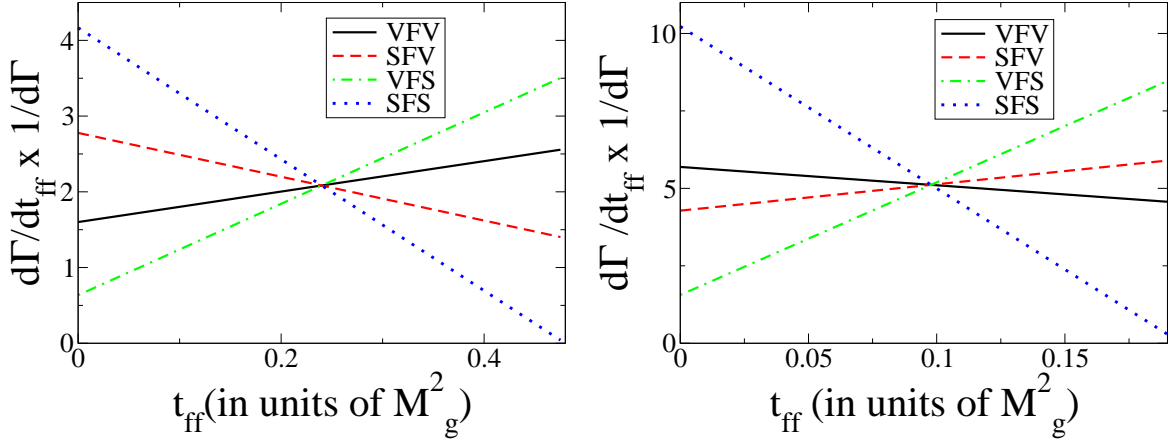

\begin{center}
\includegraphics[scale=0.3]{diff-mod-pnt1.eps}\hspace{3mm}
\includegraphics[scale=0.3]{diff-mod-pnt2.eps}
\end{center}
\caption{The resulting differential decay-width distributions corresponding to the various possibilities discussed in Table \ref{tbl:diffspin}. The mass scales are in units of $M_{g^\prime}$. The left pane corresponds to $M_Q/M_{\gamma^\prime} = 2$ and the right one has $M_Q/M_{\gamma^\prime} = 1.2$. Notice the change in the slope for the VFV and SFV models as this mass ratio is changed. } \label{fig:diffspin}
\end{figure}

\subsection{Long cascade decays and total spin determination}
\begin{figure}[ht]
\begin{center}
\includegraphics[scale=0.7]{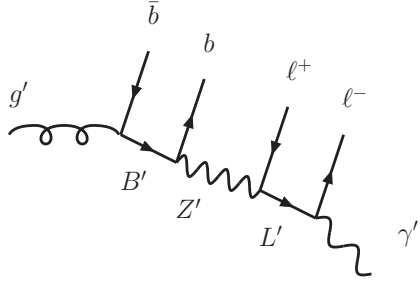}
\end{center}
\caption{A long cascade decay in scenarios with same-spin
partners. As explained in the text, a knowledge of the angular
correlations between the $b\bar{b}$ pair and dilepton
$\ell^-\ell^+$ may contain enough information to determine the spin
of all the particles in the event, including $g^\prime$, $Z^\prime$
and $\gamma^\prime$.} \label{fig:long_cascade}
\end{figure}

The cascade decay in Fig.\ref{fig:long_cascade} is an example for a decay chain where the spin of all the partners may be determined unambiguously. Suppose one measures the slope of the $b\bar{b}$ pair to be negative with
$M_{B^\prime}^2/M_{Z^\prime}^2 < 2$ and that of the dilepton pair, $\ell^-\ell^+$, to be negative with
$M_{L^\prime}^2/M_{\gamma^\prime}^2<2$ as well. Then, either all three partners, $g^\prime$, $Z^\prime$ and $\gamma^\prime$, are
vector-bosons, or all three are scalars. Hence, with a single spin measurement of the $Z^\prime$, such as described in
\cite{Barr:2004ze,Battaglia:2005zf,Smillie:2005ar,Datta:2005zs,Wang:2006hk},
one can lift this two-fold ambiguity and determine the spin of all the
particles in the event. This optimistic picture is probably
challenging in practice, however, it sharply illustrates the
utility of the slope at distinguishing between different possible spin
assignments. 

Measuring the spin in this way does not require reconstruction or a charge asymmetry. The most urgent challenge is to identify the correct pairing. Realistic simulations and experimental input will go a long way to shed some light on the issue. Also, the ability to tag the charge of the outgoing particles in the decays discussed above could be of use. A study which will quantitatively clarify how does the signal deteriorate when such information is only partially available ($b$ tagging) or altogether missing (jets) is in need. This will be helpful for assessing the prospects for spin determination at the LHC if matter partners are light enough to be present in the spectrum.

\section{Off-shell decays}
\label{sec:off-shell}
\begin{figure}[th]
\centerline{\includegraphics[scale=0.5]{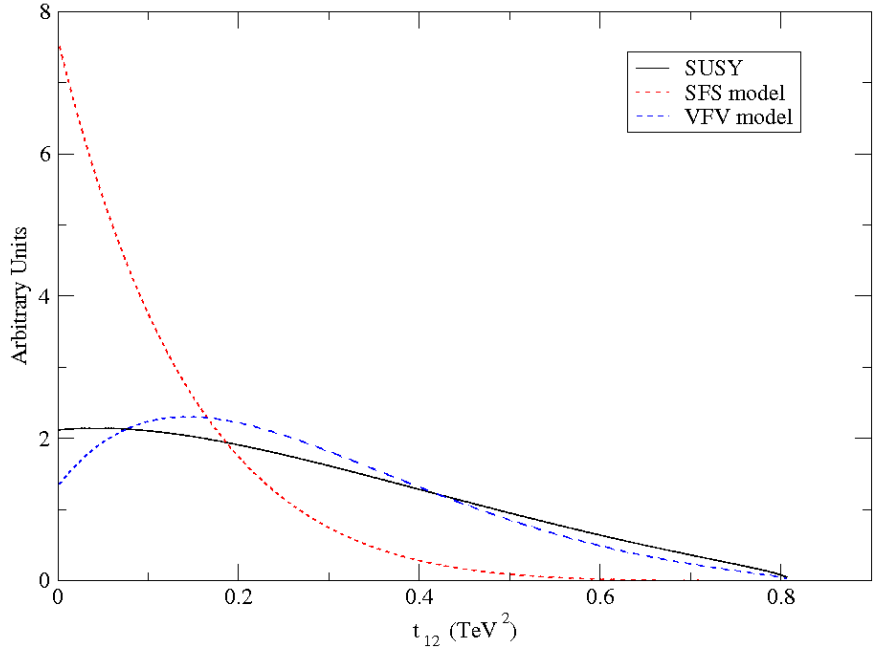}}
\vspace*{8pt}
\caption{Theoretical curves for the differential decay-width in decays of the type shown in Fig.\ref{fig:genericMab} as a function of $t_{12}=(p_1+p_2)^2$ for an off-shell intermediate state. The graphs are normalized to unit area. We contrast SUSY against two type of same-spin models. In SUSY, the intermediate particle is a scalar whereas the decaying particle and the LSP are Majorana fermions i.e. the gluino and neutralino. In VFV (SFS) scenario, the intermediate particle is a Dirac fermion and the external partners are vector-bosons (scalars).}
\protect\label{fig:off-shell}
\end{figure}

As was discussed in Ref.~\cite{Wang:2006hk} and elaborated upon in Ref.~\cite{Csaki:2007xm}, if a decay proceeds through an off-shell fermion (consider Fig.\ref{fig:intFchannel} again, but with $\bar{f}^\prime$ being off-shell), angular correlations are guaranteed because such a fermion is inevitably polarized. This can be understood by considering the fermionic propagator for a Dirac fermion,
\begin{equation}
 \Delta(p) = \frac{\slashed{p} + M}{p^2 - M^2}
\end{equation}
For $M \gg \slashed{p}$, the propagator favors one helicity structure over the other or in other words the intermediate particle is polarized.  In SUSY, the intermediate particle is a scalar and no correlations are expected. The decay's kinematics are then governed by 3-body phase-space and the resulting differential decay-width is shown in Fig.\ref{fig:off-shell}. This is contrasted with the same-spin scenario where the polarized intermediate fermion affects the distribution and causes it to deviate from the phase-space prediction. 

A simple example of such a decay is realized in scenarios where the matter partners are heavy, but the gluon partner can be produced directly (for an extreme realization of such a setup see Ref.~\cite{ArkaniHamed:2004fb}). It will then decay into neutral matter through an off-shell matter partner as in Fig. \ref{fig:intFchannel} (though, in the case of an intermediate off-shell state, one must be careful to include the crossed diagrams as well). 

On top of the experimental difficulties mentioned above, there is one additional difficulty that must be recognized. A brief look at the distributions in Fig.\ref{fig:off-shell} reveals that the real discrepancy between the different scenarios is concentrated in the low $t_{12}$ region. This corresponds to the outgoing quarks being close by in real space. It is therefore susceptible to $\Delta R$ and isolation cuts and it is not even clear that such quarks will be tagged as two different jets by the jet algorithms. These effects can carve out the entire low $t_{12}$ region and reduce the sensitivity of this observable to spin information. These effects have been included in the study of Ref.~\cite{Csaki:2007xm}, but since no jet algorithm was used, it would be important to reanalyze this channel with a more realistic simulation.

\subsection{Simulation tools to study spin correlations}

The amplitude for a single process is represented at tree level by several topologically distinct Feynman diagrams. The denominator for the amplitude corresponding to each diagram comes in the form of a product of simple poles. The numerator on the other hand takes into account the helicity structure of the diagram and can be written as the sum of several terms. To obtain an accurate simulation of the process one must square the full amplitude and include all the interference terms. Simulators such as CompHEP \cite{Boos:2004kh} and MadGraph \cite{Maltoni:2002qb} achieve exactly that albeit through slightly different routes (CompHEP offers in addition an analytic expression for the squared amplitude). 

However, when interested in events with multiple final states (more
than 4 or 5), it becomes too computationally demanding to proceed
without an approximation. If the intermediate states within each
diagram can go on-shell they will do so since such configurations
receive strong support from the pole structure of the denominator. In
this case, the interference terms between denominators with different
poles are suppressed with respect to the leading singularities. It is
then possible to approximate the full amplitude by a reduced one in
which the intermediate states are set on-shell and any interference
is neglected. This is the well-known narrow-width approximation and is
utilized by the most widely used generators: PYTHIA
\cite{Sjostrand:2006za} and Herwig \cite{Corcella:2002jc}. This
approximation of course breaks down near a threshold or when the width
becomes large compared with the mass difference (in which case the
notion of an isolated pole identified as a particle is meaningless to
begin with). It goes without saying that in the case of an off-shell
decay such an approximation is altogether invalid. 

As we just saw that the narrow-width approximation ignores interference terms between topologically distinct Feynman diagrams. 
There may still be additional interference effects associated with the
different polarizations of intermediate particles in individual
Feynman diagrams. PYTHIA and Herwig differ in the way they treat these
effects. In consequence of the narrow-width approximation the
propagator of intermediate states is reduced to the sum over
polarization and we can describe a generic matrix element involving
one propagator schematically as, 
\begin{equation}
 \langle out | prop | in \rangle = \sum_{\lambda} \langle out | \lambda \rangle \langle \lambda | in \rangle
\end{equation}
where the sum is over the different intermediate polarizations $\lambda$. The event-generator is interested in the squared amplitude which corresponds to the probability,
\begin{eqnarray}
\nonumber |\langle out |prop| in \rangle|^2 &=& \sum_{\lambda\lambda'} \langle out | \lambda \rangle \langle \lambda' | out \rangle \langle \lambda | in \rangle \langle in | \lambda' \rangle \\ &\equiv& \sum_{\lambda\lambda'}\langle in | \lambda' \rangle~ D_{\lambda\lambda'}~\langle \lambda | in \rangle 
\end{eqnarray}
In the last stage we defined the spin density matrix
$D_{\lambda\lambda'}$ which embodies all the information carried by
the numerator. Now, the spin of the initial and final state is
presumably known either because we must average or sum over it. The
question is what is being assumed about the density matrix and whether
it is to be calculated. PYTHIA simply ignores it all together setting
it to the identity matrix properly normalized. Herwig uses a clever
algorithm and computes it stochastically \cite{Collins:1987cp,
  Knowles:1988vs}. It therefore maintains full coherence as far as the
spin is concerned. It was recently extended to include heavy on-shell
vector-bosons (such as KK gluon states etc.) and the details are
described in \cite{Wang:2006hk}. 

As mentioned above the problem in using fully coherent event generator
is that it is computationally prohibitively expensive when the number
of external particles exceeds a few. Events which result in many final
states are unwieldy using this scheme. Recently, a new and useful tool called
Bridge\cite{Meade:2007js} was developed to deal with longer decay
chains by making use of the narrow-width approximation. Unlike PYTHIA,
Bridge does not ignore the density matrix altogether, but rather try to
approximate it by computing the diagonal elements,
$D_{\lambda\lambda}$, in a particular reference frame. This results
in a fairly good approximation in certain cases when one of the
elements is much larger than the others (such as in Top decay through
an intermediate $W$-boson). However, it should not be expected to
hold in general. In particular, since the density matrix is not
Lorentz invariant, while the approximation may be valid in one
reference frame it is not necessarily so in all frames. Which frame is
optimal cannot be answered in full generality. It may prove useful to try and
incorporate the algorithm for maintaining spin coherence used in
Herwig into Bridge.  

\section{Conclusion and Outlook}

We have reviewed methods of measuring the spin of new physics
particles at the LHC.  Such measurements are crucial in distinguishing
qualitatively different classes of new physics scenarios, such as
supersymmetry and Little Higgs or extra-dimensional models. 

For given masses, new physics particles with similar gauge quantum numbers but different spin
will typically have very different production rates. Therefore, combining with independent measurements of their masses,
their production cross-sections provide a quick determination of their
spin, with minimal model assumptions. However, due to
the existence of neutral stable massive particle at the end of the
decay chain in many new physics scenarios, absolute mass measurement
is challenging. Moreover, model dependent factors, such as decay
branching ratios, have to be taken into account to extract total
production rate. New method of utilizing the rate of information, such
as combining production rates in several channels and using certain
kinematical features which are sensitive to the overall mass scale,
needs to be developed. Several new approaches of measuring absolute
mass scales have been proposed recently. Their effectiveness in providing
additional information in spin measurement need to be accessed. 

We focus on direct and model independent ways of measuring spin using
angular correlation among decay products of the new physics
particles. Due to the difficulties in  fully reconstructing of the 
rest frame of the new physics particle from its decay products, we have
to use Lorentz invariant combinations 
of momenta of the observable particles, which encode the spin
information of the intermediate particles in the decay chain. We
survey the potential of extracting spin information in a large class
of different decay topologies. In each case, we present specific
observables and discuss the conditions under which spin correlations
are in principle observable. 

We considered two classes of new physics particles: partners of the
Standard Model gauge-bosons and partners of the Standard Model matter
fermions. In both cases, understanding the masses of particles and the
chiral structure of their couplings is crucial in extracting the spin
information. Details of exclusive decay chains must be well understood
so that the appropriate variables can be used and interpreted
correctly. As illustrated in the text these issues apply directly to
the gauge-boson partners. In the case of fermionic Standard Model
matter partners The couplings are always chiral when the spectrum is
non-degenerate. In the more subtle degenerate case the existence of
spin effects depends on additional parameters. In particular we
concluded that the UED model always leads to angular correlations as a
result of five-dimensional Lorentz invariance. Moreover, the slope of
the resulting distributions contains additional information regarding
the spin of the external particles. Its determination requires
knowledge of the masses of those particles. Building on these
findings, we also propose a way of measuring in principle the spin of
every 
new physics particles in the decay chain. In the case of an off-shell
decay with an intermediate fermion, angular correlations are always
present. This is an interesting direction which deserves further
detailed investigation.  

Both the existence and the interpretation of  the
spin correlation at the LHC depend strongly on the parameters of the
model under consideration. Therefore, we need to explore potentially
useful channels 
and observables as much as possible. Viable strategies of extracting
spin information at the LHC can only be finalized with  our
measurements of masses and couplings of the new physics particles.  It
is important to keep as much viable frameworks in mind as possible
while we only have incomplete information about the new physics
discovered at the LHC. 

\bibliography{ref-spinrev}

\begin{thebibliography}{10}

\bibitem{Dimopoulos:1981zb}
Savas Dimopoulos and Howard Georgi.
\newblock Softly broken supersymmetry and su(5).
\newblock {\em Nucl. Phys.}, B193:150, 1981.

\bibitem{Chung:2003fi}
D.~J.~H. Chung et~al.
\newblock The soft supersymmetry-breaking lagrangian: Theory and applications.
\newblock {\em Phys. Rept.}, 407:1--203, 2005.

\bibitem{Appelquist:2000nn}
Thomas Appelquist, Hsin-Chia Cheng, and Bogdan~A. Dobrescu.
\newblock Bounds on universal extra dimensions.
\newblock {\em Phys. Rev.}, D64:035002, 2001.

\bibitem{ArkaniHamed:2001nc}
Nima Arkani-Hamed, Andrew~G. Cohen, and Howard Georgi.
\newblock Electroweak symmetry breaking from dimensional deconstruction.
\newblock {\em Phys. Lett.}, B513:232--240, 2001.

\bibitem{Schmaltz:2005ky}
Martin Schmaltz and David Tucker-Smith.
\newblock Little higgs review.
\newblock {\em Ann. Rev. Nucl. Part. Sci.}, 55:229--270, 2005.

\bibitem{ArkaniHamed:2001ca}
Nima Arkani-Hamed, Andrew~G. Cohen, and Howard Georgi.
\newblock (de)constructing dimensions.
\newblock {\em Phys. Rev. Lett.}, 86:4757--4761, 2001.

\bibitem{Cheng:2001vd}
Hsin-Chia Cheng, Christopher~T. Hill, Stefan Pokorski, and Jing Wang.
\newblock The standard model in the latticized bulk.
\newblock {\em Phys. Rev.}, D64:065007, 2001.

\bibitem{Cheng:2001nh}
Hsin-Chia Cheng, Christopher~T. Hill, and Jing Wang.
\newblock Dynamical electroweak breaking and latticized extra dimensions.
\newblock {\em Phys. Rev.}, D64:095003, 2001.

\bibitem{Cheng:2003ju}
Hsin-Chia Cheng and Ian Low.
\newblock Tev symmetry and the little hierarchy problem.
\newblock {\em JHEP}, 09:051, 2003.

\bibitem{Cheng:2004yc}
Hsin-Chia Cheng and Ian Low.
\newblock Little hierarchy, little higgses, and a little symmetry.
\newblock {\em JHEP}, 08:061, 2004.

\bibitem{Low:2004xc}
Ian Low.
\newblock T parity and the littlest higgs.
\newblock {\em JHEP}, 10:067, 2004.

\bibitem{Cheng:2005as}
Hsin-Chia Cheng, Ian Low, and Lian-Tao Wang.
\newblock Top partners in little higgs theories with t-parity.
\newblock 2005.

\bibitem{Buckley:2007th}
Matthew~R. Buckley, Hitoshi Murayama, William Klemm, and Vikram Rentala.
\newblock Discriminating spin through quantum interference.
\newblock 2007.

\bibitem{Battaglia:2005zf}
Marco Battaglia, AseshKrishna Datta, Albert De~Roeck, Kyoungchul Kong, and
  Konstantin~T. Matchev.
\newblock Contrasting supersymmetry and universal extra dimensions at the clic
  multi-tev e+ e- collider.
\newblock {\em JHEP}, 07:033, 2005.

\bibitem{Barr:2005dz}
A.~J. Barr.
\newblock Measuring slepton spin at the lhc.
\newblock {\em JHEP}, 02:042, 2006.

\bibitem{Datta:2005vx}
AseshKrishna Datta, Gordon~L. Kane, and Manuel Toharia.
\newblock Is it susy?
\newblock 2005.

\bibitem{Meade:2006dw}
Patrick Meade and Matthew Reece.
\newblock Top partners at the lhc: Spin and mass measurement.
\newblock {\em Phys. Rev.}, D74:015010, 2006.

\bibitem{Cheng:2007xv}
Hsin-Chia Cheng, John~F. Gunion, Zhenyu Han, Guido Marandella, and Bob
  McElrath.
\newblock Mass determination in susy-like events with missing energy.
\newblock {\em JHEP}, 12:076, 2007.

\bibitem{Cho:2007qv}
Won~Sang Cho, Kiwoon Choi, Yeong~Gyun Kim, and Chan~Beom Park.
\newblock Gluino stransverse mass.
\newblock 2007.

\bibitem{Gripaios:2007is}
Ben Gripaios.
\newblock Transverse observables and mass determination at hadron colliders.
\newblock 2007.

\bibitem{Barr:2007hy}
Alan~J. Barr, Ben Gripaios, and Christopher~G. Lester.
\newblock Weighing wimps with kinks at colliders: Invisible particle mass
  measurements from endpoints.
\newblock 2007.

\bibitem{Cho:2007dh}
Won~Sang Cho, Kiwoon Choi, Yeong~Gyun Kim, and Chan~Beom Park.
\newblock Measuring superparticle masses at hadron collider using the
  transverse mass kink.
\newblock 2007.

\bibitem{Ross:2007rm}
Graham~G. Ross and Mario Serna.
\newblock Mass determination of new states at hadron colliders.
\newblock 2007.

\bibitem{Nojiri:2007pq}
M.~M. Nojiri, G.~Polesello, and D.~R. Tovey.
\newblock A hybrid method for determining susy particle masses at the lhc with
  fully identified cascade decays.
\newblock 2007.

\bibitem{Wang:2006hk}
Lian-Tao Wang and Itay Yavin.
\newblock Spin measurements in cascade decays at the lhc.
\newblock {\em JHEP}, 04:032, 2007.

\bibitem{Smillie:2006cd}
Jennifer~M. Smillie.
\newblock Spin correlations in decay chains involving w bosons.
\newblock {\em Eur. Phys. J.}, C51:933--943, 2007.

\bibitem{Hook:2008}
A.~Hook, L.~Wang, and I.~Yavin.
\newblock to be published.

\bibitem{Barr:2004ze}
A.~J. Barr.
\newblock Using lepton charge asymmetry to investigate the spin of
  supersymmetric particles at the lhc.
\newblock {\em Phys. Lett.}, B596:205--212, 2004.

\bibitem{Richardson:2001df}
Peter Richardson.
\newblock Spin correlations in monte carlo simulations.
\newblock {\em JHEP}, 11:029, 2001.

\bibitem{Alves:2006df}
Alexandre Alves, Oscar Eboli, and Tilman Plehn.
\newblock It's a gluino.
\newblock {\em Phys. Rev.}, D74:095010, 2006.

\bibitem{Smillie:2005ar}
Jennifer~M. Smillie and Bryan~R. Webber.
\newblock Distinguishing spins in supersymmetric and universal extra dimension
  models at the large hadron collider.
\newblock {\em JHEP}, 10:069, 2005.

\bibitem{Datta:2005zs}
AseshKrishna Datta, Kyoungchul Kong, and Konstantin~T. Matchev.
\newblock Discrimination of supersymmetry and universal extra dimensions at
  hadron colliders.
\newblock {\em Phys. Rev.}, D72:096006, 2005.

\bibitem{Athanasiou:2006ef}
Christiana Athanasiou, Christopher~G. Lester, Jennifer~M. Smillie, and Bryan~R.
  Webber.
\newblock Distinguishing spins in decay chains at the large hadron collider.
\newblock {\em JHEP}, 08:055, 2006.

\bibitem{Cheng:2002ab}
Hsin-Chia Cheng, Konstantin~T. Matchev, and Martin Schmaltz.
\newblock Bosonic supersymmetry? getting fooled at the lhc.
\newblock {\em Phys. Rev.}, D66:056006, 2002.

\bibitem{Kilic:2007zk}
Can Kilic, Lian-Tao Wang, and Itay Yavin.
\newblock On the existence of angular correlations in decays with heavy matter
  partners.
\newblock {\em JHEP}, 05:052, 2007.

\bibitem{Csaki:2007xm}
Csaba Csaki, Johannes Heinonen, and Maxim Perelstein.
\newblock Testing gluino spin with three-body decays.
\newblock {\em JHEP}, 10:107, 2007.

\bibitem{ArkaniHamed:2004fb}
Nima Arkani-Hamed and Savas Dimopoulos.
\newblock Supersymmetric unification without low energy supersymmetry and
  signatures for fine-tuning at the lhc.
\newblock {\em JHEP}, 06:073, 2005.

\bibitem{Boos:2004kh}
E.~Boos et~al.
\newblock Comphep 4.4: Automatic computations from lagrangians to events.
\newblock {\em Nucl. Instrum. Meth.}, A534:250--259, 2004.

\bibitem{Maltoni:2002qb}
Fabio Maltoni and Tim Stelzer.
\newblock Madevent: Automatic event generation with madgraph.
\newblock {\em JHEP}, 02:027, 2003.

\bibitem{Sjostrand:2006za}
Torbjorn Sjostrand, Stephen Mrenna, and Peter Skands.
\newblock Pythia 6.4 physics and manual.
\newblock {\em JHEP}, 05:026, 2006.

\bibitem{Corcella:2002jc}
G.~Corcella et~al.
\newblock Herwig 6.5 release note.
\newblock 2002.

\bibitem{Collins:1987cp}
John~C. Collins.
\newblock Spin correlations in monte carlo event generators.
\newblock {\em Nucl. Phys.}, B304:794, 1988.

\bibitem{Knowles:1988vs}
I.~G. Knowles.
\newblock Spin correlations in parton - parton scattering.
\newblock {\em Nucl. Phys.}, B310:571, 1988.

\bibitem{Meade:2007js}
Patrick Meade and Matthew Reece.
\newblock Bridge: Branching ratio inquiry / decay generated events.
\newblock 2007.

\end{thebibliography}
\bibliographystyle{unsrt}

\end{document}